\documentclass[pra, a4paper, preprint, showpacs,superscriptaddress]{revtex4-2}
\usepackage{amsmath,amscd,amsfonts,amssymb, amsbsy, color,bbm, bm,amsthm}
\usepackage{enumitem}
\usepackage{graphicx}{\tiny }
\usepackage{amsbsy}

\theoremstyle{plain}  
\newtheorem{thm}{Theorem}
\newtheorem{lem}{Lemma}
\newtheorem{cor}{Corollary}
\newtheorem*{pf}{Proof}
\newtheorem{rmk}{Remark}

\usepackage{longtable}

\numberwithin{equation}{section}
\begin{document}
\title{Twisted convolution quantum information channels, one-parameter semigroups and their generators}
\author{K. R. Parthasarathy}\email{krp@isid.ac.in}
\affiliation{Indian Statistical Institute, Theoretical Statistics and Mathematics Unit,Delhi Centre,
7 S.~J.~S. Sansanwal Marg, New Delhi 110 016, India} 
\date{\today}
\begin{abstract}	 
Using the tool of quantum characteristic functions of $n$-mode states in the boson Fock space $\Gamma(\mathbb{C}^n)$ we construct a semigroup of quantum information channels. This leads to a special class of one-parameter semigroups of such channels. These semigroups are concrete but their generators have unbounded operator coefficients.  These one-parameter  semigroups are also quantum dynamical semigroups and the form of the generators involve additional features which do not appear in the standard GKSL form.  A heuristic discussion of the form of these generators is included. In the wake of this analysis many open problems arise naturally.

\vskip 0.5in 

\begin{center}
	{\Large\bf\em In memory of \\ \bigskip  Robin Lyth Hudson (1940-2021) \\ and  \\  \bigskip 
		
		 Wilhelm Freiherr von Waldenfels (1932-2021). }
\end{center}

\end{abstract}

\maketitle
\newpage
\section{Notations}
\begin{itemize}
\item We shall use the Dirac notation  $\langle u\vert v\rangle$ for the scalar product  of any two elements $u$ and $v$ of a complex Hilbert space. The scalar product $\langle u\vert v\rangle$ is assumed to be  linear in $v$ and conjugate linear in $u$.  
\item Consider the   boson Fock space $\Gamma(\mathbb{C}^n)$,  over the  complex Hilbert space $\mathbb{C}^n$ of finite dimension $n$.  
\item  Fix a canonical orthonormal basis 	$\left\{ e_j, \ 1\leq j\leq n\right\}$ in  $\mathbb{C}^n$ with 
\begin{eqnarray*}
	e_j=\left( 0,0,\ldots, 0,1,0,\ldots , 0\right)^T
\end{eqnarray*}  
where '1' appears in the $j^{\rm th}$ position.  

\item At every element $\mathbf{z}\in\mathbb{C}^n$ we associate a pair of operators $a(\mathbf{z})$, $a^\dag(\mathbf{z})$,  called annihilation, creation operators, respectively in the boson Fock space $\Gamma(\mathbb{C}^n)$ and 
\begin{eqnarray*}
	W(\mathbf{z})=e^{a^\dag(\mathbf{z})-a(\mathbf{z})}
\end{eqnarray*}
denotes the  {\em Weyl operator} in $\Gamma(\mathbb{C}^n)$. 
\item Define the observables 
\begin{eqnarray}
	\label{e01}
	q(\mathbf{u})&=&\frac{a(\mathbf{u})+a^\dag(\mathbf{u})}{\sqrt{2}}  \\
	\label{e02} 
	p(\mathbf{u})&=&\frac{a(\mathbf{u})-a^\dag(\mathbf{u})}{i\,\sqrt{2}}.
\end{eqnarray}

\item Let $\mathbf{z}=\mathbf{x}+i\,\mathbf{y}=(z_1,z_2,\ldots , z_n)^T$,  $z_j=x_j+i\, y_j.$
Then  $\mathbf{x}\rightarrow q(\mathbf{x})$, 
$\mathbf{y}\rightarrow p(\mathbf{y})$ 
denote {\em position} and {\em momentum} fields obeying the commutation relations 
$$	\left[q(\mathbf{x}),\, p(\mathbf{y})\right]=i\, \mathbf{x}^T\mathbf{y}$$ 
and
$$q_j=q(e_j),\ p_j=p(e_j), \ \ j=1,2,\ldots ,n$$ 
yield the {canonical commutation relations} (CCR)~\cite{Par1992,Par10}
$$\left[q_j,\ q_k\right]=0,\ \ \left[p_j,\ p_k\right]=0,\ \ \left[q_j,\ p_k\right]= i\, \delta_{jk}.$$
\item With every $\mathbf{z}=(z_1,z_2,\ldots , z_n)^T\in \mathbb{C}^n$, $z_j=x_j+iy_j$ we associate  $2n-$dimensional phase space vector  $\bm{\xi} =(\xi_1,\xi_2,\  \ldots ,\xi_{2n})\in \mathbb{R}^{2n}$, with     $$\xi_1~=~x_1,\ \xi_2~=~y_1, \xi_3~=~x_2, \xi_4~=~y_2,\, \ldots .$$
	\end{itemize}

\noindent {\bf Correspondence between  $\mathbb{C}^n$ and the $2n-$dimensional  phase space $\mathbb{R}^{2n}$} 
	\begin{enumerate}
		\item   
  			$\mathbf{z}=(z_1,z_2,\ldots , z_n)^T  {\rm\  in}\  \mathbb{C}^n \longrightarrow  \bm{\xi}=(x_1,y_1, \  \ldots x_n,y_n) {\rm \ in\ } \mathbb{R}^{2n}.$  	        
  			  		\item  The Weyl operator $W(\mathbf{z})$ satisfies 
  		\begin{eqnarray*}
  		W(\mathbf{z})W(\mathbf{z'})&=& e^{-i\ {\rm Im}\,\langle \mathbf{z}\vert \mathbf{z'}\rangle}\, 	W(\mathbf{z+z'}) \\ 		
  	W(\mathbf{z}) &=&  \widetilde{W}(\bm{\xi}) \\ 
  	\widetilde{W}(\bm{\xi})\widetilde{W}(\bm{\xi'})&=&e^{-i\, \bm{\xi}^T\, J_{2n}\, \bm{\xi'}}\, \widetilde{W}(\bm{\xi+\xi'})  
  	\end{eqnarray*}
  where  $J_{2n}$ is a block diagonal matrix with  block of size 2:  
 	\begin{eqnarray*}
 	J_{2}=\left(\begin{array}{cc}0 & 1 \\ 1 & 0 \end{array}\right),  J_{2n}&=&J_{2}\oplus J_{2}\oplus \ldots \oplus J_2 =\left(\begin{array}{cccc}J_2 & 0 & \ldots & 0 \\ 0 & J_2 & \ldots & 0 \\ 
 	\vdots & \vdots & \ddots & \vdots\\  
 	0 & 0 & \ldots & J_2 \end{array}\right). 
 \end{eqnarray*}
When $n$ is fixed we denote $J_{2n}\equiv J$, by dropping the suffix $2n$. 
  		\end{enumerate}

\begin{rmk} 
Let $\mathbf{u},\ \mathbf{z}\in \mathbb{C}^n$ such that $\mathbf{u}\rightarrow \bm{\eta},$ $\mathbf{z}\rightarrow \bm{\xi}$, $\bm{\eta},\bm{\xi}\in\mathbb{R}^{2n}$. Then 
\begin{eqnarray}
	\label{e1}
	W(\mathbf{u})W(\mathbf{z})W(\mathbf{u})^{-1}&=& \widetilde{W}(\bm{\xi})\,e^{i\bm{\eta}^T\,J\,\bm{\xi}} \\
	 \label{e2}
	 \left[ a(\mathbf{u}), W(\mathbf{z}) \right]&=& \langle \mathbf{u}\vert \mathbf{z}\rangle\, W(\mathbf{z}) \\ 
	\label{e3}
	 \left[ a^\dag(\mathbf{u}), W(\mathbf{z}) \right]&=& \langle \mathbf{z}\vert \mathbf{u}\rangle\, W(\mathbf{z}). 
\end{eqnarray}
In other words, $W(\mathbf{z})$ is  eigen operator i.e., 
\begin{eqnarray*}
	X&\rightarrow & \left[ a(\mathbf{u}), X \right],\  
	X\rightarrow  \left[ a^\dag(\mathbf{u}), X \right] 
\end{eqnarray*}
for  $X$ in $\Gamma(\mathbb{C}^n)$.
\end{rmk}
\begin{lem}
	\label{1}
	Corresponding to the Weyl operator $W(\mathbf{z})=e^{a^{\dag}(\mathbf{z})-a(\mathbf{z})}$ we have  
	\begin{eqnarray}
		\widetilde{W}(\bm{\xi})&=& {\rm exp}\left\{i\sqrt{2}\left(\bm{\xi}^T_{\rm even}\mathbf{q}-\bm{\xi}^T_{\rm odd}\mathbf{p} \right)\right\} \nonumber \\ 
                               &=& 	{\rm exp}\left\{i\, \sqrt{2}\,\left(\sum_{j=1}^{n}\,\xi_{2j}\, q_j - \sum_{j=1}^{n}\,\xi_{2j-1}\, p_j \right)\right\} 
\end{eqnarray}
where $q_j, p_j,\ j=1,2,\ldots , n$ are position and momentum observables. 
\end{lem}
\begin{pf}
	Prove the result for one mode  and then use factorizability of $W(\mathbf{z})$ for $n$ modes.   \flushright $\square$
\end{pf}
\section{Quantum characteristic function  of a state in $\Gamma(\mathbb{C}^n)$}
With every quantum state $\rho$ in $\Gamma(\mathbb{C}^n)$ we associate a  complex-valued function  
\begin{equation}
	\label{e4}
	\hat{\rho}(\mathbf{z})={\rm Tr}\, \rho\, W(\mathbf{z}),\ \ \  \mathbf{z}\in \mathbb{C}^n
\end{equation}
called the {\em quantum  characteristic function (qcf)} of $\rho$ at $\mathbf{z}$. In terms of phase space coordinates $\bm{\xi}\in\mathbb{R}^{2n}$ we write the  qcf as 
$$f_{\rho}(\bm{\xi})={\rm Tr}\, \rho\,\widetilde{W}(\bm{\xi}).$$ 

By (\ref{e4}) of Lemma~\ref{1} we obtain 
\begin{equation}
	\label{e5}
	f_{\rho}(\bm{\xi})={\rm Tr}\, \rho\, {\rm exp}\left\{i\, \sqrt{2}\,\left(\sum_{j=1}^{n}\,\xi_{2j}\, q_j - \sum_{j=1}^{n}\,\xi_{2j-1}\, p_j \right)\right\}
	\end{equation}
Writing 
$$	f_{\rho,\bm{\xi}}(t)=	f_{\rho}(t\bm{\xi}),$$ 
for $t\in\mathbb{R}$ and $\bm{\xi}$ fixed in $\mathbb{R}^{2n}$ we present the following Lemma. 
\begin{lem}
	\label{3}
\noindent (i) $f_{\rho,\bm{\xi}}(t)$ as a function of $t$ is a classical characteristic function of the distribution of the observable 
\begin{equation}
	\label{e5a}
	p(\bm{\xi})=\sqrt{2}\,\left(\sum_{j=1}^{n}\,\xi_{2j}\, q_j - \sum_{j=1}^{n}\,\xi_{2j-1}\, p_j\right)
\end{equation}
for every $\bm{\xi}$ in $\mathbb{R}^{2n}.$ \\ 
\noindent (ii) $f_{\rho}(\xi_1,0,\xi_3,0,\ldots , \xi_{2n-1}, 0)$ as a function of  $\xi_1,\xi_3,\ldots , \xi_{2n-1}$ is the classical characteristic function of the joint probability distribution of the commuting sequence of the observables $-\sqrt{2} \left(p_1,p_2,\ldots, p_n\right)$ in the state $\rho$. \\
\noindent(iii) Similarly, $f_{\rho}(0,\xi_2,0,\xi_4,0,\ldots , 0, \xi_{2n})$ is the classical characteristic function of the joint probability distribution of the commuting sequence of the observables $\sqrt{2} \left(q_1,q_2,\ldots, q_n\right)$ in the state $\rho$.
\end{lem}
\begin{pf}
	Immediate from (\ref{e5}). \flushright$\square$
\end{pf}
  
We proceed to present quantum Bochner theorem~\cite{Par1992,Par10} in phase space language.
\begin{thm} 
	A continuous complex valued function $f(\bm{\xi})$, $\bm{\xi}\in\mathbb{R}^{2n}$ such that $f(\mathbf{0})=1$ is the quantum characteristic function  $f_{\rho}(\bm{\xi})$ of a state $\rho$ in $\Gamma(\mathbb{C}^n)$ if and only if the kernel $K_{f}(\bm{\xi},\bm{\xi}')=f_\rho(\bm{\xi}'-\bm{\xi})\, e^{i\,\bm{\xi}^T\,J\,\bm{\xi}'}$ is positive definite; such a $\rho$ is unique. 
\end{thm}
\begin{pf}
	Immediate from the relation 
	$${\rm Im}\langle\mathbf{z}\vert \mathbf{z'}\rangle=\bm{\xi}^T\, J\, \bm{\xi'}$$
	where $\mathbf{z},\  \mathbf{z'}$ in $\mathbb{C}^n$ correspond to $\bm{\xi,\ \xi'}$ in phase space $\mathbb{R}^{2n}.$  \flushright $\square$
\end{pf}
\begin{cor}
	If $\rho$ is any state in $\Gamma(\mathbb{C}^n)$ and $\varphi(\mathbf{\xi})$ is the characteristic function of a probability distribution in $\mathbb{R}^{2n}$ then 
	\begin{equation}
		\left(f_{\rho}\,\varphi \right)(\bm{\xi})=f_{\rho}(\bm{\xi})\,\varphi(\bm{\xi})=f_{\rho'}(\bm{\xi})
	\end{equation}
for a unique state $\rho'$ determined by $\rho$ and $\varphi$.  
\end{cor}
\begin{pf}
	Immediate.
	 \flushright  $\square$
\end{pf}
\noindent Now we elaborate the  Corollary~1.  

Let $\mathbf{u,\ z}\in\mathbb{C}^{n}$, $\mathbf{u}\longrightarrow\bm{\eta}$,  $\mathbf{z}\longrightarrow\bm{\xi}$    in phase space $\mathbb{R}^{2n}$. 
Let $\rho$ be any state in $\Gamma(\mathbb{C}^n)$. Then 
\begin{eqnarray*}
	{\rm Tr}\, W(\mathbf{u})^{-1}\,\rho\, W(\mathbf{u})\,W(\mathbf{z})&=&e^{2i\, {\rm Im}\langle\mathbf{u}\vert\mathbf{z}\rangle}\,{\rm Tr} \rho W(\mathbf{z}) \\
	&=& e^{2i\,\bm{\eta}^T J \bm{\xi}}\,f_{\rho}(\bm{\xi})\, 
\end{eqnarray*}
Let now $\bm{\eta}$ be an $\mathbb{R}^{2n}$-valued random variable with the probability distribution $\mu$ on the borel $\sigma$-algebra of $\mathbb{R}^{2n}$.
Then 
\begin{eqnarray*}
\mathbb{E}_{\mu}	{\rm Tr}\, W(\mathbf{u})^{-1}\,\rho\, W(\mathbf{u})\,W(\mathbf{z})&=&f_{\rho}(\bm{\xi})   \mathbb{E}_{\mu}\, e^{i\,\bm{\eta}^T\,2J\, \bm{\xi}}  \\
	&=&  f_{\rho}(\bm{\xi})\, \hat{\mu}(2J\bm{\xi})
\end{eqnarray*}
 where $\mathbf{u}=\mathbf{u}(\bm{\eta})$ is the random variable in $\mathbb{C}^n$ corresponding to the random variable $\eta$ in $\mathbb{R}^{2n}$ amd $\hat{\mu}$ is the characteristic function of $\mu$ in $\mathbb{R}^{2n}.$  If we write $\varphi(\bm{\xi})=\hat{\mu}(2J\bm{\xi})$ where $\varphi$ is again a classical characteristic function and $f_{\rho}\,\varphi=f_{\rho'}$ for the state $\rho'$ in Corollary~1. In view of this relation we call the map 
 $$T(\varphi)\rho=\rho'$$
 where 
 $$f_{\rho'}(\bm{\xi})=f_{\rho}(\bm{\xi})\,\varphi(\bm{\xi})$$
 the {\em convolution channel}. For every trace class operator $\rho$ in $\Gamma(\mathbb{C}^n)$ write again '
 $$f_{\rho}(\bm{\xi})=	{\rm Tr}\, \rho\,\widetilde{W}(\bm{\xi}).$$
Denote by $\Phi$ the linear space of all such functions defined on the phase space. Denote by $\Phi_B$ the space of all continuous complex valued functions $f(\bm{\xi})$ on $\mathbb{R}^{2n}$ which satisfy the following conditions: 
\begin{eqnarray*}
	f(\bm{0})=1,\ f(-\bm{\xi})=\bar{f(\bm{\xi})}
\end{eqnarray*}
and the hermitian kernel 
$K_f(\bm{\xi,\ \eta})=f(\bm{\eta-\xi})\, e^{i\, \bm{\xi}^T\,J\, \bm{\eta}}$,  $\bm{\xi},\ \bm{\eta}\in \mathbb{R}^{2n}$ is positive definite. By the quantum Bochner theorem $f_{\rho}(\bm{\xi})=	{\rm Tr}\, \rho\,\widetilde{W}(\bm{\xi})$ $\forall\ \bm{\xi}$ for a unique state $\rho$ in $\Gamma(\mathbb{C}^n).$ We call any element of the space  $\Phi_B$  a quantum Bochner function.  Then $\Phi_B\subset\Phi$ is a convex set. From now on a quantum Bochner function is simply termed as Bochner function. 

From the discussions above it follows that for any Bochner function $f$, i.e., an element of $\Phi_B$, and any classical characteristic function $\varphi(\bm{\xi})$ of a probability distribution in 
 $\mathbb{R}^{2n},$ the function $f\, \varphi$ is also a Bochner function.  
 
  We write 
 $$\widetilde{T}(\varphi)\,f_{\rho}=f_{\rho}\, \varphi=f_{\rho'}$$
 the convolution channel in the space of {\em Bochner functions}.

We summarize: 
\begin{itemize}
 \item  Every Bochner function $f$ on $\mathbb{R}^{2n}$ is of the form $f_{\rho}$, the qcf of a state $\rho$ in $\Gamma(\mathbb{C}^n).$
\item All Bochner functions on $\mathbb{R}^{2n}$ constitute a convex set. 
\item For any classical characteristic function $\varphi$ of a probability distribution on $\mathbb{R}^{2n}$ the map 
$\widetilde{T}(\varphi)\,f=f\,\varphi,\ \ f\in\Phi_B,$ where $\Phi_B$ is the space of all (quantum) Bochner functions, is called a convolution channel. 
\item  $\widetilde{T}$ is an affine linear map on the convex set $\Phi_B$. 
\item The map  
$$T(\varphi)\,\rho=\rho'$$
with 
$$f_{\rho'}=f_{\rho}\,\varphi$$
is called a quantum convolution channel. 
\item The map $T(\varphi)$ is an affine linear map on the convex set of all states in $\Gamma(\mathbb{C}^n)$.     
\end{itemize}

\begin{thm}
	Let $\mu$ be a probability distribution on $\mathbb{R}^{2n}$ with characteristic function $\hat{\mu}$ and let $\varphi(\bm{\xi})=\hat{\mu}(2J\bm{\xi})$. Let $\rho$ be any state in 
	$\Gamma(\mathbb{C}^n)$. Consider 
	$$\widetilde{T}(\varphi)\,f_{\rho}=f_{\rho}\,\varphi=f_{\rho'}.$$
	Then 
	$$\rho'=\int_{\mathbb{R}^{2n}} \widetilde{W}(\bm{\eta})\,\rho\, \widetilde{W}(\bm{\eta})^{-1}\, \mu(d\bm{\eta})$$
	where the operator integral in the right hand side is in the weak sense  and $\widetilde{W}(\bm{\eta})$ is the Weyl operator in the phase space variable.  \hskip 4.5in $\square$
\end{thm}
\begin{rmk}
	Since the  integrand is a displacement of $\rho$ and $\mu$ is a distribution in $\mathbb{R}^{2n}$ it is natural to call $\rho'$ the convolution $\rho\star\mu$. Thus the nomenclature 
	\begin{eqnarray*}
		\widetilde{T}(\varphi)&&: {\rm\  Convolution\  channel\  in\ } \Phi_B \\ 
		T(\varphi):&& {\rm\  Convolution\ channel\  in\  the\  state\  space.} 
		\end{eqnarray*}
\end{rmk}
\section{Twisted convolution channels}

We start with two real $2n\times 2n$ matrices $A$, $M$ where $M$ is positive definite. For any state $\rho$ in $\Gamma(\mathbb{C}^n)$ with qcf $f_\rho(\bm{\xi}),$ $\bm{\xi}\in\mathbb{R}^{2n}$ define 
\begin{equation}
	\label{e7}
		\widetilde{T}(\varphi)\,f_\rho(\bm{\xi})=e^{-\frac{1}{2}\,\bm{\xi}^T M\bm{\xi}}\,f_{\rho}(A\bm{\xi}). 
	\end{equation} 
Now we ask under what conditions the right hand side of (\ref{e7}) is a quantum Bochner function i.e., the kernel 
\begin{equation}
	\label{e8}
	K_f(\bm{\xi,\, \eta})=e^{i\,\bm{\xi}^T J\bm{\eta}}\,e^{-\frac{1}{2}\,(\bm{\eta-\xi})^T\,M\,(\bm{\eta-\xi})}\,f_{\rho}(A\left(\bm{\eta-\xi})\right) 
\end{equation} 
is positive definite. To this end we note that the right hand side of (\ref{e8}) is a product of 3 terms: 
\begin{enumerate}
	\item \hskip 0.3in $e^{i\,(A\bm{\xi})^T J (A\bm{\eta})}\,f_{\rho}(A\bm{\eta}-A{\bm\xi}).$
	\item \hskip 0.3in $e^{i\,\bm{\xi}^T J\,\bm{\eta}-i\,(A\bm{\xi})^T J (A\bm{\eta})+\,\bm{\xi}^T M\bm{\eta} }.$
	\item \hskip 0.3in $e^{-\frac{1}{2}\,\bm{\xi}^T\,M\,\bm{\xi}}\,e^{-\frac{1}{2}\,\bm{\eta}^T\,M\,\bm{\eta}}.$
\end{enumerate}
It follows that the first factor is a positive definite Kernel since $f_{\rho}$ is a qcf  (quantum Bochner theorem). The third term is of the form $g(\bm{\xi})\,g(\bm{\eta})$ where $g(\xi)=e^{-\frac{1}{2}\,\bm{\xi}^T\,M\,\bm{\xi}}$ is real and hence corresponds to a positive definite kernel. The second term will be positive definite if its exponent 
$$\bm{\xi}^T\, \left(M +i\,(J- A^T\, J\, A)\right)\bm{\eta})$$ 
is {\em conditionally} positive definite~\cite{KRP&Schmidt1972}. This holds if and only if 
$$M +i\,(J- A^T\, J\, A)\geq 0$$
in the sense of left hand side being a positive semidefinite matrix. 

Thus we have obtained the following theorem. 

\begin{thm}
	\label{thm7}
	Let $A, \, M$ be $2n\times 2n$ real matrices such that $M$ is positive definite and the matrix inequality  
	\begin{equation}
		\label{e9}
		M +i\,(J- A^T\, J\, A)\geq 0
	\end{equation} 
holds. Then for any state $\rho$ in $\Gamma(\mathbb{C}^n)$ there exists a unique state $\rho'$ in $\Gamma(\mathbb{C}^n)$ such that 
$$f_{\rho'}(\bm{\xi})=e^{-\frac{1}{2}\,\bm{\xi}^T\,M\,\bm{\xi}}\, f_\rho(A\,\bm{\xi}),\ \ \forall \ \bm{\xi}\in\mathbb{R}^{2n}.$$
	\end{thm}
The map $f_\rho\longrightarrow f_{\rho'}$ is affine linear on the convex set $\Phi_B$ of all Bochner functions. \hskip 0.1in $\square$ \\ 
We write 
\begin{eqnarray}
	\label{e10}
	\rho'&=& T(M,A)\,\rho \nonumber \\ 
	f_{\rho'}&=& \widetilde{T}(M,\,A)\,f_{\rho}.
\end{eqnarray}
\begin{rmk}
	The map  $T(M,A): \rho\rightarrow \rho',$   with $\rho, \rho'$ in $\Gamma(\mathbb{C}^n)$ is a quantum information channel, while  $\widetilde{T}(M,\,A):f_{\rho}\rightarrow f_{\rho'},$\ with $f_\rho,\ f_{\rho'}\in \Phi_B$  is a $\Phi_B$-space channel. 
\end{rmk}

\begin{cor}
	Let $\varphi$ be the classical characteristic function of a probability distribution in $\mathbb{R}^{2n}$ and let $T(\varphi)$ be the associated convolution channel and $\widetilde{T}(\varphi)$ the corresponding $\Phi_B$-channel. Let $T(M,A),\ \widetilde{T}(M,A)$ be as in Theorem~\ref{thm7}. Then 
	\begin{eqnarray*}
		T(\varphi, M, A)&=& T(\varphi)\, T(M,A) \\ 
		\widetilde{T}(\varphi, M, A)&=& \widetilde{T}(\varphi)\, \widetilde{T}(M,A)  
	\end{eqnarray*}
obtained by composition are respectively quantum information and $\Phi_B$-space channels. 
\end{cor}
\noindent \textbf{Definition:}  The channel $T(\varphi, M, A)$ is called a {\em twisted convolution information channel} and $\widetilde{T}(\varphi, M, A)$, the corresponding {\em twisted convolution $\Phi_B$-channel} associated  with the triple $(\varphi,\,M\, A).$ 
\subsection{Examples of $(M\,A)$ satisfying (\ref{e9}) of Theorem~\ref{thm7}.}
\begin{itemize}
\item Note that $J-A^T\,J\,A$ is real skew symmetric and $i\,(J-A^T\,J\,A)$ being hermitian, has real eigenvalues. Thus $J-A^T\,J\,A$ has  all its eigenvalues purely imaginary. If $\lambda$ is an eigenvalue of $J-A^T\,J\,A$, so is $\bar{\lambda}.$ Thus for all $ \lambda_1\geq\lambda_2\geq \ldots\geq \lambda_n$ such that the eigenvalues of the hermitian matrix $i(J-A^T\,J\,A)$ are $\pm\lambda_1,\,\pm\lambda_2,\ldots,\pm\lambda_n$. Thus for all positive definite $M'$, the matrix $\lambda_1\, I_{2n}+M'=M$ together with $A$ satisfies the inequality (\ref{e9}). 

In particular, when $A$ is symplectic (i.e.,\ $\lambda_1, \lambda_2,\ldots,\lambda_n=0$) any $M\geq 0$ together with $A$ satisfies (\ref{e9}).

\item Information Channel $T(\varphi,M,A)$ and $\Phi_B$-channel $\widetilde{T}(\varphi, M, A)$: 
\begin{eqnarray*}
\rho_{\rm in}=\rho &\longrightarrow &\framebox{$T(\varphi,M,A)$}  \longrightarrow \rho_{\rm out}=\rho'\ \ \ \  {\rm (state\ space)}  \\
f_\rho&\longrightarrow &\framebox{$\widetilde{T}(\varphi,M,A)$}  \longrightarrow f_{\rho'}\ \  \ \  (\Phi_B\ {\rm space)} \\ 
f_{\rho'}(\bm{\xi})&=&f_{\rho}(A\bm{\xi})e^{-\frac{1}{2}\,\bm{\xi}^T\,M\,\bm{\xi}}\,\varphi(\bm{\xi}). \\ 
\end{eqnarray*}
	\item For fixed $\bm{\xi}$ it is seen that  $f_{\rho',\bm{\xi}}(t), \ t\in\mathbb{R}$ is the characteristic function of the distribution of the observable $p(\bm{\xi})$ (see (\ref{e5a})) in the state $\rho_{\rm out}=\rho'$. This is the product of three characteristic functions: (i) $f_{\rho,\, A\bm{\xi}}$ (ii) the normal characteristic function $e^{-\frac{t^2}{2}\bm{\xi}^TM\bm{\xi}}$ and 
	(iii) $\varphi_{\bm{\xi}}(t)=\varphi(t\,\bm{\xi})$ on $\mathbb{R}$ (and hence the convolution of the corresponding distributions on $\mathbb{R}$. See Lemma~\ref{3}.)
	\item Let $\Pi_1\,\bm{\xi}= (0,\xi_2,0,\xi_4,0,\ldots , 0,\xi_{2n}).$ Then  the  characteristic function of the joint  distribution of  $\sqrt{2}\,\mathbf{q}= \sqrt{2}\,\left(q_1,q_2,\ldots, q_n\right)$ in $\mathbb{R}^n$ in the state $\rho_{\rm out}$ is the product of two characteristic functions in $\mathbb{R}^n$ namely, 
	$$f_{\rho}(A\,\Pi_1\,\bm{\xi})\, e^{-\frac{1}{2}\,\bm{\xi}^T\,\Pi_1^T\, M\, \Pi_1\,\bm{\xi}}$$ 
	and $\phi(\Pi_1\,\bm{\xi})$ as functions of $\xi_2,\xi_4,\ldots , \xi_{2n}.$  
	\item A similar result holds for the joint distribution of $-\sqrt{2}\mathbf{p}=-\sqrt{2} \left(p_1,p_2,\ldots, p_n\right)$ in the state $\rho_{\rm out}$ with $\Pi_1$ replaced by $\Pi_2$ where 
	 $\Pi_2\,\bm{\xi}= (\xi_1,0,\xi_3,0,\ldots , \xi_{2n-1},0).$
	\item Let $\rho_{in}=\rho$ be a gaussian state. Then 
	$$f_{\rho}(\bm{\xi})=e^{i\,\bm{\lambda}^T\,\bm{\xi}-\bm{\xi}^T\,K\,\bm{\xi}}$$
	for some $\bm{\lambda}\in\mathbb{R}^{2n}$ and $K$ is the covariance matrix of $\rho.$ 
	Then 
	\begin{eqnarray*}
		\left[\widetilde{T}(\varphi,M,A)\, f_\rho\right](\bm{\xi})=\, e^{\bm{\lambda}^T\,A\,\bm{\xi}-\bm{\xi}^T\,A^T\,K\,A\,\bm{\xi}-\frac{1}{2}\bm{\xi}^T\,M\,\bm{\xi}}\, \varphi(\bm{\xi})=f_{\rho'}(\bm{\xi})
	\end{eqnarray*}
where $\rho'=\rho_{\rm out}$ is the convolution of a gaussian state with transformed covariance matrix $A^T\,K\,A+\frac{1}{2}\,M$ passed through $\widetilde{T}(\phi).$ 

If $\varphi$ is also a gaussian (normal) classical characteristic function, it follows that $\rho_{\rm out}$ is a gaussian state. In other words $T(\phi,M,A)$ is a gaussian channel in the sense that $\rho_{\rm out}$ is gaussian whenever $\rho_{\rm in}$ is gaussian.  
 \end{itemize}
\section{Composition of twisted convolution channels}
Consider $T(\varphi_j,M_j,A_j)$ with $j=1,2.$ Let $\rho$ be a state in $\Gamma(\mathbb{C}^n)$ with qcf $f_{\rho}(\bm{\xi}),\ \bm{\xi}\in \mathbb{R}^{2n}.$ Then 
$$\left[\widetilde{T}(\varphi_1,M_1,A_1)\, f_\rho\right](\bm{\xi})=f_\rho(A_1\,\bm{\xi})\, e^{-\frac{1}{2}\bm{\xi}^T M \bm{\xi}}\,\varphi_1(\bm{\xi})=f_{\rho'}(\bm{\xi}).$$ 
We then obtain 
\begin{eqnarray*}
\left[\widetilde{T}(\varphi_2,M_2,A_2)\, f_{\rho'}\right](\bm{\xi})&=&f_\rho(A_1\,A_2\,\bm{\xi})\, e^{-\frac{1}{2}\bm{\xi}^T A_2^T M_1 A_2\bm{\xi}}\, \varphi_1(A_2\,\bm{\xi}) 
\,  e^{-\frac{1}{2}\bm{\xi}^T\,M_2\,\bm{\xi}}\, \varphi_2(\bm{\xi}) \\ 
&=& \left[\widetilde{T}(\varphi_2(\varphi_1\cdot A_2),M_2+A_2^T\,M_1\,A_2,A_1\,A_2)\, f_\rho\right](\bm{\xi}).
\end{eqnarray*}
Thus 
\begin{equation}
	\label{e11}
\framebox{$T(\varphi_2,M_2,A_2)\,T(\varphi_1,M_1,A_1)=T(\varphi_2(\varphi_1\cdot A_2), \ M_2+A_2^T\,M_1\,A_2,A_1\,A_2)$}.
\end{equation}
\noindent \textbf{Note}: Since $\varphi_1$ is a classical characteristic function so is $\varphi_1\cdot A_2$, with $A_2$ being a linear transformation $\bm{\xi}\longrightarrow A_2\bm{\xi}.$ Since $\varphi_2$ is also a characteristic function so is the product  $\varphi_2~(\varphi_1~\cdot~A_2).$ 

From our derivation of the composition rule it follows that the pair  $\left(M_2+A_2^T\,M_1\,A_2,\, A_1\,A_2\right)$ satisfies the requirement (\ref{e9}) from the same property for 
$\left(M_j,\, A_j\right),\ j=1,2.$ It is also possible to prove the same by elementary algebra. 

Note that (\ref{e11}) holds when $T$ is replaced by $\widetilde{T}$ for twisted convolution $\Phi_B$-channels. 

\begin{thm}
	Twisted convolution channels form a semigroup with the composition law (\ref{e11}). 
\end{thm} 
\section{One-parameter semigroups of twisted convolution channels}
The composition rule for twisted convolution channels in (\ref{e11}) suggests the search for one-parameter semigroups of the form 
\begin{eqnarray*}
	T_t&=& T(\varphi_t,M_t,A_t), \ t\geq 0\\ 
	T_0&=& T(1,0,I_{2n})={\rm Identity\ channel} \\ 
	T_s\,T_t&=& T_{s+t}\ \forall\ s\geq 0,\ t\geq 0.  
\end{eqnarray*}
In terms of the parameters $(\varphi,M,A)$, with $\varphi$ being classical characteristic function in $\mathbb{R}^{2n}$, the set $(M,A)$ of matrices obeying (\ref{e9}),  the composition rule (\ref{e11}) yields the following: \\ 
\begin{eqnarray}
	\label{ei}
	 A_s\,A_t&=& A_{s+t} \ \ \forall\ s\geq 0,\ t\geq 0,\ A_0=I_{2n} \\ 
	 \label{eii}
	 M_{t+s}&=&M_t+A_t^{T}\,M_s\, A_t \\ 
	\label{eiii}
	 \varphi_{t+s}&=&\varphi_t  (\varphi_s\cdot A_t). 
\end{eqnarray}
\begin{itemize}
\item Assume continuity in $s,\, t$. Then it follows from (\ref{ei}) that $A_t=e^{t\,A}$ for a fixed $2n\times 2n$ matrix $A$. From (\ref{eii}) it is seen that $M_0=0$. 
\item Assuming differentiability in $t,\, s$ and differentiating at $s=0$ we obtain 
$$\dot{M}_t=e^{t\,A^T}\, \dot{M}_0\, e^{t\,A}.$$ 
\item Define $N=\dot{M}_0$. Then 
\begin{eqnarray}
	\label{e12}
	M_t=\int_{0}^{t}\, e^{\tau\,A^T}\, N\, e^{\tau\, A}\, d\tau.
\end{eqnarray} 
Since $M_t\geq0$, $M_t/t\geq 0,\ \ \forall\ t>0$, taking limit $t\downarrow\, 0$ in (\ref{e12}) we conclude that $N$ is positive semidefinite. 
\item From (\ref{e9}) we have 
\begin{eqnarray*}
	M_t+ i\, \left(J-e^{t\,A^T}\, J\, e^{t\,A}\right)\geq 0.
\end{eqnarray*} 
Dividing both sides by $t$ and letting $t\downarrow\, 0$, we get 
\begin{eqnarray}
	\label{e13}
	N+ i\, \left(A^T\, J +J\, A\right)\geq 0.
\end{eqnarray} 
 \item Starting from (\ref{e13}), left and right multiplication by $e^{\tau\, A^T},\ e^{\tau\, A}$ respectively and integrating with respect to $t=\tau$ in the interval $[0,\tau]$, we see that 
 $$M_t-i\, \left(e^{t\, A^T}\, J\, e^{t\, A}\right)\geq 0$$
 and the same holds when $i$ is changed to $-i$. In other words $\left(M_t,\, e^{t\,A}\right)$ obeys (\ref{e9}).   
\item  Now we turn our attention to the parameter $\varphi_t$ in the channel $T_t$ of (\ref{eiii}). Replace $t$ by $t-\frac{1}{m},\ s=\frac{1}{m}$ where $m$ is an integer and $\frac{1}{m}\leq t$ with $t>0$ fixed. Then 
\begin{eqnarray*}
	\varphi_t&=& \varphi_{t-\frac{1}{m}} (\varphi_{\frac{1}{m}}\cdot A_{t-\frac{1}{m}}) \\ 
	&=&  \varphi_{t-\frac{1}{m}}\left[\varphi_{\frac{1}{m}}\cdot A_{t-\frac{1}{m}}\right]\cdot \left[\varphi_{\frac{1}{m}}\cdot A_{t-\frac{2}{m}}\right] \\ 
	&& \vdots \hskip 0.6in \vdots\\ 
	&=& \varphi_{t-\frac{mt}{m}}\prod_{r=1}^{[mt]}\,\varphi_{\frac{1}{m}}\cdot A_{t-\frac{r}{m}}
\end{eqnarray*}
by successive approximation of (\ref{eiii}) with $t$ replaced by $t-\frac{r}{m}$ and $s$ by $\frac{r}{m}$, \ $r=1,2,\ldots ,\, [mt]$ with $[mt]=$ integral part of $t, \ m=1,2,\ldots. $ 

Thus we have expressed $\varphi$ as a limit of a triangular array product of uniformly infinitesimal characteristic functions~\cite{Kol1954}. Thanks to the L{\'e}vy-Khintchine theory, $\varphi_t$ is an infinitely divisible characteristic function and 
\begin{eqnarray*}
	\varphi_t=e^{\psi_t},\ \ t\geq 0,\ \ \psi_0=1
\end{eqnarray*} 
where $\psi_t$ has the L{\'e}vy-Khintchine representation. 
\item Now the multiplication equation (\ref{eiii}) becomes the additive equation:
\begin{eqnarray*}
	\psi_{t+s}=\psi_t+\psi_s\cdot A_t, \ A_t=e^{t\,A}.
\end{eqnarray*} 
\item The map $t\mapsto \psi_t$ is continuous. Assuming differentiability in $t$ for each fixed $\bm{\xi}\in\mathbb{R}^{2n}$ and differentiating at $s=0$ we get 
\begin{eqnarray*}
	\dot{\psi}_{t}=\dot{\psi}_{0}\cdot A_t. 
\end{eqnarray*}
\item Put $\gamma=\dot{\psi}_{0}$, $\gamma(\bm{\xi})=\frac{d}{dt}\left.\psi_t(\bm{\xi})\right\vert_{t=0}\ \ \forall\ \ \bm{\xi}\in\mathbb{R}^{2n}.$
Solving for $\psi_t$ we get 
\begin{eqnarray*}
	\psi_t(\bm{\xi})=\int_0^t\,\gamma\left(e^{\tau\,A}\right)\, d\tau
\end{eqnarray*}
\end{itemize}
The  L{\'e}vy-Khintchine functions $\psi_t$ are known to be conditionally positive in $\bm{\xi}$. If we assume $\gamma\left(\bm{\xi}\right)$ to be continuous in $\bm{\xi}$ then it follows that $\gamma$ is also conditionally positive and $\gamma$ is also a L{\'e}vy-Khintchine function in $\mathbb{R}^{2n}$. Furthermore $\gamma\left(\bm{\xi}\right)$ admits the representation 
\begin{eqnarray}
	\label{e15} 
   \gamma\left(\bm{\xi}\right)&=& -\, \bm{\xi}^T\, B\, \bm{\xi}- i \bm{\lambda}^T\, \bm{\xi} + \int_{\mathbb{R}^{2n}\backslash \{0\}} \,  \left(e^{i\,\bm{\eta}^T\bm{\xi}}-1-\frac{i\bm{\eta}^T\bm{\xi}}{1+\vert\bm{\eta}\vert^2}\right)\, dF(\bm{\eta})
\end{eqnarray}
where $B\geq 0$ and  $F$ is a  L{\'e}vy measure with following properties: 
\begin{eqnarray}
	\label{e16}
	(a)\hskip 0.2in  &&  F\left(\left.\left\{\bm{\eta}\,\right\vert\, \vert\bm{\eta}\vert>\varepsilon\right\}\right)<\infty,\ \ \forall\ \varepsilon>0 \nonumber \\
	(b)\hskip 0.2in  && \int_{0<\vert\bm{\eta}\vert<\varepsilon}\,  \vert\bm{\eta}\vert^2\,  dF(\bm{\eta}) <\infty \ {\rm \ for \ some\ } \varepsilon>0.
	\end{eqnarray} 
From the expression for $\gamma$ in (\ref{e15}), we can drop the quadratic form $\bm{\xi}^T B \bm{\xi}$ by clubbing $B$ with $N$ and then replacing $N$ by $\mathcal{N}=N+B$. This leads us to a basic example of twisted convolution channels. We state this in the following theorem.  
\begin{thm}
	\label{thm10}
	Let $A$ and $\mathcal{N}$ be $2n\times 2n$ real matrices where $\mathcal{N}$ is a positive definite matrix obeying the inequality 
	$$\mathcal{N}+i\,(A^T\,J+J\, A)\geq 0.$$  
Let $\gamma(\bm{\xi})$ be the L{\'e}vy-Khintchine function defined by  
$$ \gamma\left(\bm{\xi})\right)=  \int_{\mathbb{R}^{2n}\backslash \{0\}} \,  \left(e^{i\,\bm{\eta}^T\bm{\xi}}-1-\frac{i\bm{\eta}^T\bm{\xi}}{1+\vert\bm{\eta}\vert^2}\right)\, dF(\bm{\eta})$$
 where $F$ is any fixed   L{\'e}vy measure in $\mathbb{R}^{2n}\backslash \{0\}$ satisfying conditions (a), (b) in (\ref{e16}). Define the matrices 
 $$M_t=\int_{0}^t\, e^{\tau\, A^T}\, \mathcal{N}\, e^{\tau\, A}\, d\tau, \ t\geq 0$$
 and the infinitely divisible characteristic functions $\varphi_t$ with the L{\'e}vy-Khintchine representation: 
 $$\varphi_t(\bm{\xi})={\rm exp}\left\{\int_0^t\,\gamma\left(e^{\tau\, A}\,\bm{\xi}\right)\, d\tau.\right\} $$
 Then the twisted convolution channels 
 $$T_t=T(\varphi_t,M_t, e^{t\, A}),\  t\geq 0$$ are well defined and 
 $$T_s\, T_t=T_{s+t}$$ 
 for all $s,t\geq 0$.  In particular, for any state $\rho_0$ in $\Gamma(\mathbb{C}^n)$ the qcf of the state $\rho_t=T_t\, \rho_0$ is given by 
 \begin{equation}
 	\label{e17}
 f_{\rho_t}(\bm{\xi})=f_{\rho_0}\left(e^{tA\,\bm{\xi}}\right)\, e^{-\frac{1}{2}\bm{\xi}^T\,M_t\bm{\xi}}\varphi_t(\bm{\xi}),\ \ t\geq 0,\ \bm{\xi}\in\mathbb{R}^{2n}
 \end{equation}
where $f_{\rho_0}$ is the qcf of $\rho_0$.     \hskip 4in  $\square$
\end{thm}
In the example of the concrete one-parameter semigroup $\{T_t\}$ of twisted convolution channel it is natural to investigate the dynamical behaviour of the state $\rho_t=T_t\,\rho_0$ for a given initial state $\rho_0$ as $t$ increases. Since $\rho_t$ is completely determined by its qcf 
\begin{equation}
	\label{e18}
	\Psi(t,\bm{\xi})=f_{\rho_t}(\bm{\xi}),\ t\geq 0, \ \bm{\xi}\in\mathbb{R}^{2n}
\end{equation}
 given by (\ref{e17}) varies as $t$ increases. To this end, we observe that $\Psi(t,\bm{\xi})$ is differentiable in $\bm{\xi}$ \ $\forall \ \ t>0$ if 
  $	\Psi(0,\bm{\xi})=f_{\rho}(\bm{\xi})$ is once differentiable in $\bm{\xi}$. This is immediate from the expression in the right hand side of (\ref{e17}). 
  
  By differentiating in $t$ at $t=0$, note that 
  $$\overset{\lim}{t\downarrow 0}\, \frac{\Psi(t,\bm{\xi})-\Psi(0,\bm{\xi})}{t}=\bm{\xi}^T\,A^T\,\nabla_{\bm{\xi}}\,f_{\rho_0}+\left(-\frac{1}{2}\,\bm{\xi}^T\mathcal{N}\bm{\xi}+\gamma(\bm{\xi})\,f_{\rho_0}(\bm{\xi})\right).$$ 
  Denote 
  $$\left(\frac{\partial}{\partial\,t}\right)_+ \Psi(t,\bm{\xi})= \overset{\lim}{h\downarrow 0}\, \frac{\Psi(t+h,\bm{\xi})-\Psi(t,\bm{\xi})}{h}.$$
  It follows from the semigroup property of $\{\widetilde{T}_t\}$ that 
$$\left(\frac{\partial}{\partial\,t}\right)_+ \Psi(t,\bm{\xi})= \bm{\xi}^TA^T\, \nabla_{\bm{\xi}}\Psi(t,\bm{\xi}) + V(\bm{\xi})\, \Psi(t,\bm{\xi})$$
where 
\begin{eqnarray}
	\label{evxi}
V(\bm{\xi})&=& -\bm{\xi}^T\,\mathcal{N}\bm{\xi}+\gamma(\bm{\xi}) \nonumber \\ 
&=& -\bm{\xi}^T\,\mathcal{N}\bm{\xi}+\int_{\mathbb{R}^{2n}\backslash \{0\}}\, \left(e^{i\,\bm{\eta}^T\bm{\xi}}-1-\frac{i\,\bm{\eta}^T\bm{\xi}}{1+\vert\bm{\eta}\vert^2}\right)\, dF(\bm{\eta}).
\end{eqnarray}
It is interesting to note that $e^{V(\bm{\xi})}$ is an infinitely divisible characteristic function with  gaussian factor 
$e^{-\frac{1}{2}\, \bm{\xi}^T\,\mathcal{N}\bm{\xi}}$ and a L{\'e}vy factor 
$${\rm exp}\left\{\int_{\mathbb{R}^{2n}\backslash \{0\}} \left( e^{i\,\bm{\eta}^T\bm{\xi}}-1-\frac{i\,\bm{\eta}^T\bm{\xi}}{1+\vert\bm{\eta}\vert^2}\right)\, dF(\bm{\eta})\right\}$$
with L{\'e}vy measure $F$. 

We may say that the first order differential operator 
\begin{equation}
	\label{e19}
	\mathcal{D}=\bm{\xi}^T\,A^T\,\nabla_{\bm{\xi}} + V(\bm{\xi}) 
\end{equation}
is the dynamical generator of $\{\widetilde{T}_t\}$ and the equation 
$$\left(\frac{\partial}{\partial\,t}\right)_+ \Psi(t,\bm{\xi})= \left(\bm{\xi}^TA^T\, \nabla_{\bm{\xi}} + V(\bm{\xi})\right)\, \Psi(t,\bm{\xi}).$$

We proceed to give a heuristic derivation of the generator of the channel semigroup $\{T_t,\ t\geq 0\}.$ 

We create a linear space containing the states on $\Gamma(\mathbb{C}^n)$ in which linear operations may be done comfortably.
 To this end  we choose the space of all trace class operators on $\Gamma(\mathbb{C}^n)$, {\em viz.,} $\mathcal{B}_1\left(\Gamma(\mathbb{C}^n)\right)$. 
 Denoting an arbitrary element by $\rho$  write 
	\begin{eqnarray*}
		\hat{\rho}(\mathbf{z})={\rm Tr}\, \rho\, W(\mathbf{z})=f_{\rho}(\bm{\xi})
	\end{eqnarray*} 
where $\mathbf{z}=\left(z_1,z_2,\ldots , z_n\right),\ \ z_j=x_j+i\,y_j,$\, $\bm{\xi}=(x_1,y_1,x_2,y_2,\ldots x_n,y_n)^T$ in $\mathbb{R}^{2n}.$ 
For convenience in calculations we write 
$$f_{\rho}(\bm{\xi})=f(\rho,\bm{\xi}).$$
For translating the action of the generator $\mathcal{D}$ (see (\ref{e19})) on qcf  $f(\rho,\bm{\xi})$ we have to translate (i) the multiplication operations  
$$f(\rho,\bm{\xi})\longrightarrow \xi_j\, f(\rho,\bm{\xi})$$ 
and (ii) the differentiations 
$$f(\rho,\bm{\xi})\rightarrow \frac{\partial}{\partial\,\xi_j}f(\rho,\bm{\xi})$$
into operations on $\rho$. This involves unbounded operators in $\mathcal{B}_1\left(\Gamma(\mathbb{C}^n)\right).$ To this end we shall use a heuristic approach. To this end we shall use a heuristic approach. Define 
$$L_X\,Y=X\,Y, \ R_X\,Y=Y\,X$$
where the operator $Y$ varies and $X$ remains fixed. $L_X,\ R_X$ are called {\em left} and {\em right} multiplications by $X$. 
\begin{itemize}
\item Recall the commutation rules  (see (\ref{e2}),(\ref{e3})): 
\begin{eqnarray*}
\left[a_j,\, W(\textbf{z})\right]&=&z_j\, W(\mathbf{z}) \\ 
\left[a^\dag_j,\, W(\textbf{z})\right]&=&\bar{z}_j\, W(\mathbf{z}), \ \ \ \ 1\leq j\leq n.  
\end{eqnarray*} 
\item Using $q_j=\frac{a_j+a_j^\dag}{\sqrt{2}},\ \ p_j=\frac{a_j-a_j^\dag}{i\,\sqrt{2}}$   (see (\ref{e01}), (\ref{e02})) deduce 
\begin{eqnarray*}
	\left[q_j,\, W(\textbf{z})\right]&=&\sqrt{2}\, x_j\, W(\mathbf{z}) \\ 
	\left[p_j,\, W(\textbf{z})\right]&=&\sqrt{2}\, y_j\, W(\mathbf{z}),\ \ 1\leq j\leq n.  
\end{eqnarray*} 
\end{itemize}
 Let us consider the one-mode case where 
\begin{eqnarray}
	W(z)&=& W(x+iy) \nonumber \\ 
	&=& W(x) \, W(iy)\, e^{i\, x\, y}\nonumber \\ 
	&=& e^{-i\,\sqrt{2}\, p}\, e^{-i\,\sqrt{2}\,y\, q}\,  e^{i\, x\, y} \nonumber \\ 
	\label{e20}
	\frac{\partial}{\partial\,x}\, W(z)&=& -i\,\sqrt{2}\,p\, W(z) +i\, y\, W(z) \\ 
		\label{e21}
	\frac{\partial}{\partial\,y}\, W(z)&=& i\,\sqrt{2}\,\, W(z)\, q +i\, x\, W(z).  
\end{eqnarray}
For $\rho$ in $\mathcal{B}_1\left(\Gamma(\mathbb{C})\right)$ we obtain 
\begin{eqnarray*}
	{\rm Tr}\, \rho\, 	\left[q,\, W(z)\right] &=& \sqrt{2}\,	{\rm Tr}\, \rho\, W(z)= \sqrt{2}\,x\, f(\rho, \bm{\xi}) \\ 
	&=&   \sqrt{2}\,\xi_1\, f(\rho, \bm{\xi}) \nonumber \\ 
	{\rm Tr}\, \rho\, 	\left(q\, W(z)-W(z)\, q\right) &=& {\rm Tr}\, \left(R_q-L_q\right)\rho\,W(z) \\ 
	&=& f\left(\left(R_q-L_q\right)\rho,\ \bm{\xi}\right)  
\end{eqnarray*}
provided $\left(R_q-L_q\right)\rho$ is defined as a trace class operator. 

Thus 
\begin{equation}
	\label{e22}
	\xi_1\, f(\rho,\bm{\xi})=f\left(\frac{R_q-L_q}{\sqrt{2}}(\rho),\bm{\xi}\right).
\end{equation}
Same argument with $\rho$ and $p$ yields 
\begin{equation}
	\label{e23}
	\xi_2\, f(\rho,\bm{\xi})=f\left(\frac{R_p-L_p}{\sqrt{2}}(\rho),\bm{\xi}\right).
\end{equation}
Equations (\ref{e22}), (\ref{e23}) show that multiplication by phase variables $\xi_1$, $\xi_2$  become respectively the unbounded operators 
$$\rho\longrightarrow \frac{R_q-L_q}{\sqrt{2}},\ \ \rho\longrightarrow \frac{R_p-L_p}{\sqrt{2}}$$
which commute. 

We shall now use (\ref{e20}) and (\ref{e23}):
\begin{eqnarray}
	\frac{\partial}{\partial\,\xi_1}\,f(\rho, \bm{\xi})&=& -i\,\sqrt{2}\,	{\rm Tr}\, \rho\,p\, W(z) +i\, y\, {\rm Tr}\, \rho\, W(z) \nonumber \\
	&=&  -i\,\sqrt{2}\, 	{\rm Tr}\, R_p\,\rho\, W(z) +i\, \xi_2\, f(\rho,\bm{\xi}) \nonumber \\ 
	&=& i\, f(-\sqrt{2}\,R_p\,\rho,\bm{\xi}) +i\, 
	f\left(\frac{R_p-L_p}{\sqrt{2}}(\rho),\,\bm{\xi}\right)\nonumber \\
	\label{e24} 
{\rm Thus}\ \ \ \ \frac{\partial}{\partial\,\xi_1}\,f(\rho, \xi)	&=&  f\left(\frac{R_p+L_p}{i\,\sqrt{2}}(\rho),\,\bm{\xi}\right) 
\end{eqnarray}
A similar computation using (\ref{e21}) and (\ref{e22}) leads to 
\begin{equation}
	\label{e24} 
\frac{\partial}{\partial\,\xi_2}\,f(\rho, \bm{\xi})	=  f\left(\frac{R_q+L_q}{-i\,\sqrt{2}}(\rho),\,\bm{\xi}\right). 
\end{equation}
Note that the actions $R_p+L_p$ and $R_q+L_q$ commute.

\begin{itemize}
\item  Action of $\left(\begin{array}{cc}\xi_1 & \xi_2\end{array}\right)\, A^T\, \left(\begin{array}{c}\frac{\partial}{\partial\xi_1} \\ \frac{\partial}{\partial\xi_2}\end{array}\right)$ on $f(\rho,\bm{\xi})$  turns out to be 
$$\rho\rightarrow \left(\begin{array}{cc}\frac{R_q-L_q}{\sqrt{2}}, & \frac{R_p-L_p}{\sqrt{2}}\end{array}\right)\, A^T\,J_2^T \left(\begin{array}{c}\frac{R_q+L_q}{-i\,\sqrt{2}}, \\ \frac{R_p+L_p}{i\,\sqrt{2}}\end{array}\right)(\rho).$$
\item  Action of $V(\bm{\xi})=V(\xi_1,\xi_2)$ on $f(\rho,\bm{\xi})$ becomes 
$$\rho\rightarrow V\left(\frac{R_q-L_q}{\sqrt{2}},  \frac{R_p-L_p}{\sqrt{2}}\right)\,\left(\rho\right).$$
 \end{itemize}
Thus the differential equation  for the states $\rho_t,\ t\geq 0$ (see  Theorem~\ref{thm10}) becomes 
\begin{eqnarray*}
	\left(\frac{d}{dt}\right)_+\rho_t&=& \left\{\frac{1}{2}\left(\begin{array}{cc} R_q-L_q, & R_p-L_p \end{array}\right)\, A^T\,J_2^T\,\left(\begin{array}{c} i\, (R_q+L_q) \\ -i\,(R_p+L_p)\end{array}\right)+
	V\left(\frac{R_q-L_q}{\sqrt{2}},  \frac{R_p-L_p}{\sqrt{2}}\right)\right\}\,\rho_t.
\end{eqnarray*}
In other words the {\em generator} $\mathcal{L}$ of $\{T_t, t\geq 0\}$ has the form 
\begin{eqnarray}
	\mathcal{L}\,\rho_t&=& \left\{\frac{1}{2}\left(\begin{array}{cc}R_q-L_q, & R_p-L_p\end{array}\right)\, A^T\,J_2^T\,\left(\begin{array}{c} i\, (R_q+L_q) \\ -i\,(R_p+L_p) \end{array}\right)+
	V\left(\frac{R_q-L_q}{\sqrt{2}},  \frac{R_p-L_p}{\sqrt{2}}\right)\right\}\,\rho_t.\nonumber \\
\end{eqnarray} 

Turning to the differentiation of the function $V(\bm{\xi})$ we see that it is just the quadratic form $-\frac{1}{2}\,\bm{\xi}^T\, \mathcal{N}\,\bm{\xi}$  when the L{\'e}vy measure $F$ is 0 (see (\ref{evxi})) and $	V\left(\frac{R_q-L_q}{\sqrt{2}},  \frac{R_p-L_p}{\sqrt{2}}\right)(\rho)$ is easily understood to be  
$$-\frac{1}{2}\, \left(\begin{array}{cc}\frac{R_q-L_q}{\sqrt{2}}, & \frac{R_p-L_p}{\sqrt{2}}\end{array}\right)\, \mathcal{N}\, \left(\begin{array}{c} \frac{R_q-L_q}{\sqrt{2}}\\ \frac{R_p-L_p}{\sqrt{2}} \end{array}\right)(\rho).$$ 
It is made of sums of the form $X\,\rho\, Y$, $X\,Y\,\rho$, $\rho\,X\, Y$, where $X,\ Y$ vary over $\pm\,q,\ \pm\,p$. However $	V\left(\frac{R_q-L_q}{\sqrt{2}},  \frac{R_p-L_p}{\sqrt{2}}\right)$ needs a deeper analysis  when $F$ is not 0. This needs to be developed. 

Going to the case of $n$-modes we can write down the form of the generator $\mathcal{L}$. Note that $\left\{R_{q_j}-L_{q_j}.R_{p_j}-L_{p_j},\ j=1,2,\ldots, n\right\}$ is a commutative family. Similarly 
 $\{R_{q_j}~+~L_{q_j}$, $R_{p_j}~+~L_{p_j},\ j~=~1,2,\ldots, n\}$ is another commutative family. Then the generator of $\{T_t, t\geq 0\}$ has the form 
 \begin{eqnarray*}
 	\mathcal{L}&=& \frac{1}{2}\left(\begin{array}{ccccc}R_{q_1}-L_{q_1}, & R_{p_1}-L_{p_1} &  \ldots , &  R_{q_n}-L_{q_n}, & R_{p_n}-L_{p_n}  \end{array}\right)\, A^T\,J_{2n}^T\,\left(\begin{array}{c} i\,(R_{q_1}+L_{q_1}) \\  -i\,(R_{p_1}+L_{p_1}) \\ \vdots \\  i\,(R_{q_n}+L_{q_n})\\ -i\,(R_{p_n}+L_{p_n})  \end{array}\right) \nonumber \\ 
 && + \, V\left(\frac{R_{q_1}-L_{q_1}}{\sqrt{2}}, \,\frac{R_{p_1}+L_{p_1}}{\sqrt{2}}, \ldots ,\, \frac{R_{q_n}-L_{q_n}}{\sqrt{2}}, \,\frac{R_{p_n}+L_{p_n}}{\sqrt{2}}\right). \\
 	&& \hskip 0.3in \rho\longrightarrow\, \mathcal{L}\,\rho,  \ \ \ \ 	{\rm with\  }\ \ \left(\frac{d}{dt}\right)_+\rho_t=\mathcal{L}\, \rho_t. 
 \end{eqnarray*}
\section{Conclusions} 

The central construction in this paper is a class of one parameter semigroups of quantum information channels $\{T_t\}$  and their {\em avataras}  $\{\widetilde{T}_t\}$ in the phase space description. There are two fundamental matrices $(A,\, M)$ involved in this construction: \begin{itemize}
	\item $A$ is an arbitrary $2n\times 2n$ real matrix \item $M$ is a real $2n\times 2n$ positive definite matrix. 
	\item $A$ and $M$ satisfy the matrix inequality $$M+i\,\left(J-A^T\,J\, A\right)\geq 0$$
	where   
	\begin{eqnarray*}
		  J&=&J_{2}\oplus J_{2}\oplus \ldots \oplus J_2 =\left(\begin{array}{cccc}J_2 & 0 & \ldots & 0 \\ 0 & J_2 & \ldots & 0 \\ 
			\vdots & \vdots & \ddots & \vdots\\  
			0 & 0 & \ldots & J_2 \end{array}\right)
	\end{eqnarray*} 
is a $2n\times 2n$ matrix containing  $2\times 2$ matrix blocks  $J_{2}=\left(\begin{array}{cc}0 & 1 \\ 1 & 0 \end{array}\right).$ 
\end{itemize}
There arise some open problems in the wake of this analysis: 
\begin{enumerate}
	\item What is the most general pair $(A,\, M)$ of $2n\times 2n$ real matrices so that the twisted convolution information channel $T(\phi,\, M,\, A)$ of Sec.~3, with $\varphi$ being a classical characteristic function of a probability distribution in $\mathbb{R}^{2n}$, can be defined?   
	\item There exists a well understood theory of characteristic functions of probability measures in infinite dimensional Hilbert spaces. There is also an extensive theory of weak convergence and L{\'e}vy-Khintchine theory of infinitely divisible distributions in a Hilbert space~\cite{KRP_book}. Using this theory it should be possible to extend the present work to the case of infinite number of modes in a boson Fock space. 
	\item Theorem~5 of the paper is central. It gives a formula for $f_{\rho_t}=T_t\, f_{\rho_0}$. The infinitely divisible characteristic functions $\{\varphi_t,t\geq 0\}$ are all knit together in a single jump process with independent increments carrying values in $\mathbb{R}^{2n}$. The associated L{\'e}vy measure of this process is given by $\widetilde{F}$ where 
	$$\widetilde{F}([0,t])=\int_{0}^{t}\,\left(F\,e^{-\tau\,A^T}\right)\, d\tau,\ \ \ t\geq 0.$$
	It would be interesting to understand this process in terms of the observables related to the dilation of $\{T_t\}.$
	\item The distinguishing feature of our example is the appearance of the L{\'e}vy jump process in the channel and hence the quantum dynamical semigroup. When the  L{\'e}vy measure is zero the generator of the dynamical semigroup is quadratic in $\bf{p},\ \bf{q},$  which is equivalent to  quadratic generator involving $\bf{a},\, \bf{a}^\dag .$ In this case the channel turns out to be gaussian. It may be interesting to cast the same GKSL form and compare it with the result of Teretenkov~\cite{Ter}. Non-Markovian semigroups are also featured in the review article by Teretenkov~\cite{Ter}. These non-Markovian semigroups seem to require lot of clarification in terms of the underlying processes. It is desirable to cast the bosonic generators in the phase space language and analyse the existence of semigroups.   
\end{enumerate}

\section*{Acknowledgements} 
I thank Professor A.~R.~Usha Devi of Department of Physics, Bangalore University, Begaluru for going through my algebraic scribblings and transforming them into this readable manuscript. I also express my appreciation for several useful conversations on the subject over the phone  in the course of the preparation of this manuscript. My thanks to Mrs.~Shyamala Parthasarathy for her patience in handling  e-communications  between Delhi and Bengaluru. 

Support from the Delhi Center of the Indian Statistical Institute in the form of an Emeritus Position is gratefully acknowledged

\end{document}